\documentclass[aps,prl,twocolumn,amsmath,amssymb,superscriptaddress]{revtex4-2}
\usepackage[utf8]{inputenc}
\usepackage{latexsym}
\usepackage{amsfonts}
\usepackage{amssymb}
\usepackage{amsmath}
\usepackage{array}
\usepackage{dcolumn}
\usepackage{longtable}
\usepackage[dvipdfmx]{graphicx}
\usepackage{color}
\usepackage[normalem]{ulem}
\usepackage{xcolor}
\usepackage{soul,xcolor}
\usepackage{float}
\usepackage{multirow}
\usepackage{siunitx}

\usepackage{url}
\usepackage{hyperref}
\hypersetup{
    colorlinks,
    citecolor=blue,
    filecolor=orange,
    linkcolor=cyan,
    urlcolor=green}
\usepackage{orcidlink}

\usepackage{dcolumn}
\newcolumntype{d}{D{.}{.}{-1}}
\newcolumntype{f}[1]{D{.}{.}{#1}}
\newcommand{\eg}{{\textit{e.g., }}}
\newcommand{\ie}{i.e., }

\begin{document}

\title{Few-electron highly charged muonic Ar atoms verified by electronic $K$ x rays}

\author{T. Okumura\orcidlink{https://orcid.org/0000-0002-3037-6573}}\email{tokumura@tmu.ac.jp}
\affiliation{Department of Chemistry, Tokyo Metropolitan University, Hachioji, Tokyo 192-0397, Japan}
\author{T. Azuma\orcidlink{https://orcid.org/0000-0002-6416-1212}}
\email{toshiyuki-azuma@riken.jp}
\affiliation{Atomic, Molecular and Optical Physics Laboratory, RIKEN, Wako 351-0198, Japan}
\author{D. A. Bennett}
\affiliation{National Institute of Standards and Technology, Boulder, Colorado 80305, USA}
\author{W. B. Doriese}
\affiliation{National Institute of Standards and Technology, Boulder, Colorado 80305, USA}
\author{M. S. Durkin} 
\affiliation{Department of Physics, University of Colorado, Boulder, CO, 80309, USA}
\author{J. W. Fowler} 
\affiliation{National Institute of Standards and Technology, Boulder, Colorado 80305, USA}
\author{J. D. Gard}
\affiliation{Department of Physics, University of Colorado, Boulder, CO, 80309, USA}
\author{T. Hashimoto}
\affiliation{RIKEN Cluster for Pioneering Research, RIKEN, Wako, 351-0198, Japan}
\affiliation{RIKEN Nishina Center, RIKEN, Wako 351-0198, Japan}
\author{R. Hayakawa}
\affiliation{International Center for Quantum-field Measurement Systems for Studies of the
Universe and Particles (QUP), High Energy Accelerator Research Organization (KEK), Tsukuba, Ibaraki 305-0801, Japan}
\author{Y. Ichinohe}
\affiliation{RIKEN Nishina Center, RIKEN, Wako 351-0198, Japan}
\author{P. Indelicato\orcidlink{https://orcid.org/0000-0003-4668-8958}}
\affiliation{Laboratoire Kastler Brossel, Sorbonne Universit\'{e}, CNRS, ENS-PSL Research University, Coll\`{e}ge de France, Case 74, 
 4, place Jussieu, 75005 Paris, France}
\author{T. Isobe}
\affiliation{RIKEN Nishina Center, RIKEN, Wako 351-0198, Japan}
\author{S. Kanda}
\affiliation{High Energy Accelerator Research Organization (KEK), Tsukuba, Ibaraki 305-0801, Japan}
\author{D. Kato}
\affiliation{National Institute for Fusion Science (NIFS), Toki, Gifu 509-5292, Japan}
\affiliation{Interdisciplinary Graduate School of Engineering Sciences, Kyushu University, Fukuoka 816-8580, Japan}
\author{M. Katsuragawa}
\affiliation{Kavli IPMU (WPI), The University of Tokyo, Kashiwa, Chiba 277-8583, Japan}
\author{N. Kawamura}
\affiliation{High Energy Accelerator Research Organization (KEK), Tsukuba, Ibaraki 305-0801, Japan}
\author{Y. Kino}
\affiliation{Department of Chemistry, Tohoku University, Sendai, Miyagi 980-8578, Japan}
\author{N. Kominato}
\affiliation{Department of Physics, Rikkyo University, Tokyo 171-8501, Japan}
\author{Y. Miyake}
\affiliation{High Energy Accelerator Research Organization (KEK), Tsukuba, Ibaraki 305-0801, Japan}
\author{K. M. Morgan}
\affiliation{National Institute of Standards and Technology, Boulder, Colorado 80305, USA}
\affiliation{Department of Physics, University of Colorado, Boulder, CO, 80309, USA}
\author{H. Noda}
\affiliation{Department of Astronomy, Tohoku University, Sendai, Miyagi 980-8578, Japan}
\author{G. C. O'Neil}
\affiliation{National Institute of Standards and Technology, Boulder, Colorado 80305, USA}
\author{S. Okada}\email{sokada@isc.chubu.ac.jp}
\affiliation{Department of Mathematical and Physical Sciences, Chubu University, Kasugai, Aichi 487-8501, Japan}
\affiliation{National Institute for Fusion Science (NIFS), Toki, Gifu 509-5292, Japan}
\author{K. Okutsu}
\affiliation{Department of Chemistry, Tohoku University, Sendai, Miyagi 980-8578, Japan}
\author{N. Paul\orcidlink{https://orcid.org/0000-0003-4469-780X}}
\affiliation{Laboratoire Kastler Brossel, Sorbonne Universit\'{e}, CNRS, ENS-PSL Research University, Coll\`{e}ge de France, Case 74, 
 4, place Jussieu, 75005 Paris, France}
\author{C. D. Reintsema}
\affiliation{National Institute of Standards and Technology, Boulder, Colorado 80305, USA}
\author{T. Sato}
\affiliation{Department of Physics, Meiji University, Kawaski, Kanagawa 214-8571, Japan}
\author{D. R. Schmidt}
\affiliation{National Institute of Standards and Technology, Boulder, Colorado 80305, USA}
\author{K. Shimomura}
\affiliation{High Energy Accelerator Research Organization (KEK), Tsukuba, Ibaraki 305-0801, Japan}
\author{P. Strasser}
\affiliation{High Energy Accelerator Research Organization (KEK), Tsukuba, Ibaraki 305-0801, Japan}
\author{D. S. Swetz}
\affiliation{National Institute of Standards and Technology, Boulder, Colorado 80305, USA}
\author{T. Takahashi\orcidlink{https://orcid.org/0000-0001-6305-3909}}
\affiliation{Kavli IPMU (WPI), The University of Tokyo, Kashiwa, Chiba 277-8583, Japan}
\author{S. Takeda}
\affiliation{Kavli IPMU (WPI), The University of Tokyo, Kashiwa, Chiba 277-8583, Japan}
\author{S. Takeshita}
\affiliation{High Energy Accelerator Research Organization (KEK), Tsukuba, Ibaraki 305-0801, Japan}
\author{M. Tampo}
\affiliation{High Energy Accelerator Research Organization (KEK), Tsukuba, Ibaraki 305-0801, Japan}
\author{H. Tatsuno}
\affiliation{Department of Physics, Tokyo Metropolitan University, Tokyo 192-0397, Japan}
\author{K. T\H{o}k\'{e}si}
\affiliation{HUN-REN Institute for Nuclear Research (ATOMKI), 4026 Debrecen, Hungary}
\author{X. M. Tong}
\affiliation{Center for Computational Sciences, University of Tsukuba, Tsukuba, Ibaraki 305-8573, Japan}
\author{Y. Toyama}
\affiliation{Center for Muon Science and Technology, Chubu University, Kasugai, Aichi, 487-8501, Japan}
\author{J. N. Ullom}
\affiliation{National Institute of Standards and Technology, Boulder, Colorado 80305, USA}
\affiliation{Department of Physics, University of Colorado, Boulder, CO, 80309, USA}
\author{S. Watanabe}
\address{Department of Space Astronomy and Astrophysics, Institute of Space and Astronautical Science (ISAS), Japan Aerospace Exploration Agency (JAXA), Sagamihara, Kanagawa 252-5210, Japan}
\author{S. Yamada}
\affiliation{Department of Physics, Rikkyo University, Tokyo 171-8501, Japan}
\author{T. Yamashita}
\affiliation{Department of Chemistry, Tohoku University, Sendai, Miyagi 980-8578, Japan}
%

\begin{abstract}
{\it Electronic} $K$ x rays emitted by muonic Ar atoms in the gas phase were observed using a superconducting transition-edge-sensor microcalorimeter. The high-precision energy spectra provided a clear signature of the presence of muonic atoms accompanied by a few electrons, which have never been observed before. 
One-, two-, and three-electron bound, \ie H-like, He-like, and Li-like, muonic Ar atoms were identified from electronic $K$ x rays and hyper-satellite  $K$ x rays. 
These $K$ x rays are emitted after the charge transfer process by the collisions with surrounding Ar atoms. 
With the aid of theoretical calculations, 
we confirmed that the peak positions are consistent with the x-ray energies from highly charged Cl ions, and the intensities reflecting deexcitation dynamics were successfully understood by taking into account the interaction between the muon and bound electrons.
\end{abstract}

\maketitle

\begin{figure}[htb]
	\begin{center}
		\includegraphics[keepaspectratio=true,,width=0.9\linewidth]{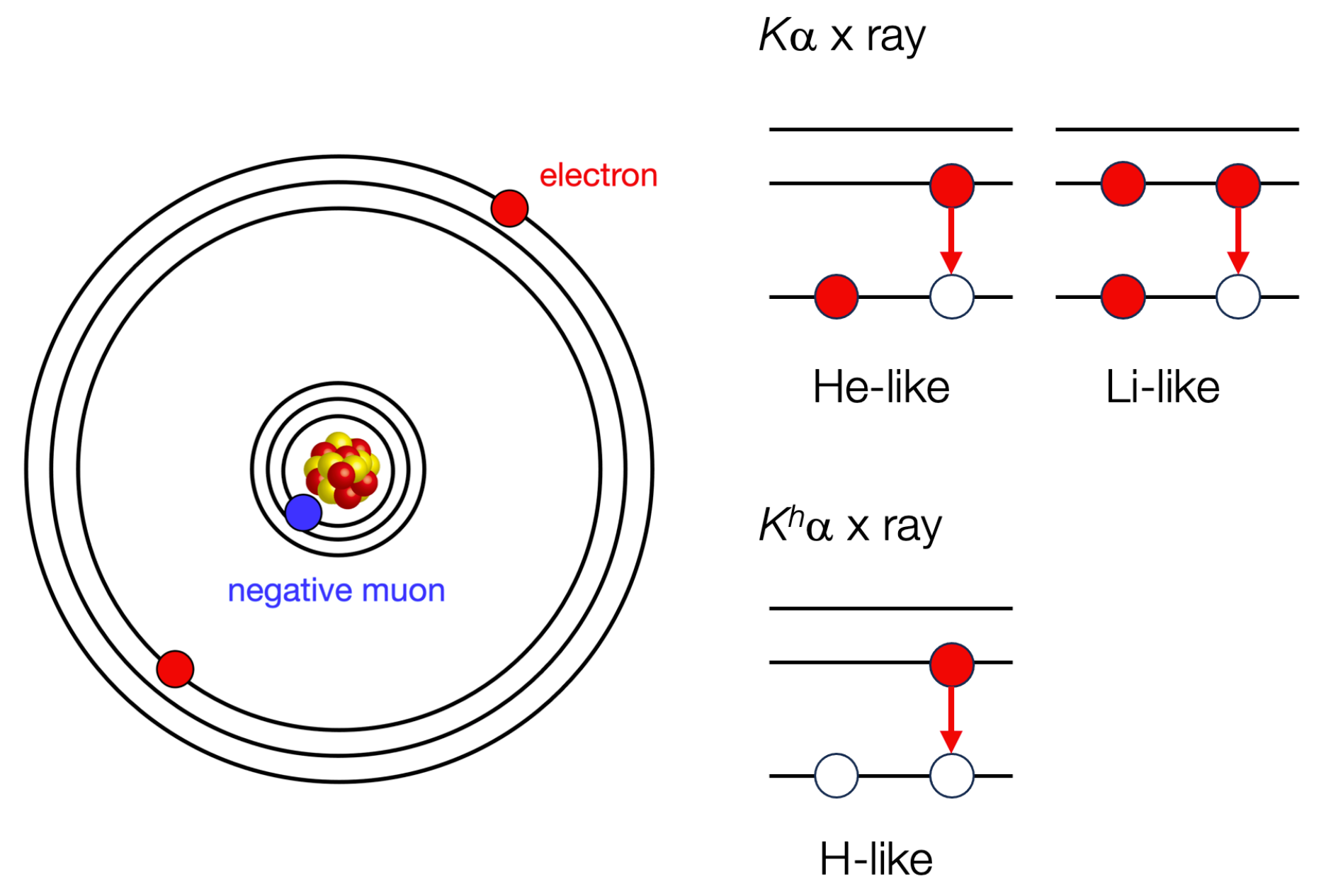}
	\end{center}%
	\caption{ \label{fig:scheme} 
	Schematic drawing of highly charged (He-like) muonic atoms with two bound electrons (left) and electronic configuration for $K\alpha$ x-ray and hyper-satellite $K^h\alpha$ x-ray emissions (right).}
\end{figure}

Highly charged ions (HCI) play a key role in many fields of science, such as fundamental physics, nuclear fusion plasmas, surface science, and astronomy.
Their structure and behavior have been intensively studied via spectroscopy of the emitted characteristic x rays using various types of ion sources, e.g., electron beam ion traps \cite{beiersdorfer2003,bbkl2007,mlcn2008,kmmu2014,hell2020}, electron-cyclotron resonance ion sources \cite{asgl2012,mssa2018,mbpt2020,mpsd2023} and highly-charged ions accelerators \cite{gsbb2005,gths2018,lbds2024}. 
Few-electron HCIs prepared in storage rings or ion traps have contributed to benchmarking atomic structure calculations and  verification of quantum electrodynamics (QED) in strong electromagnetic fields (see \eg \cite{ind2019,akr2023} for recent reviews). 
HCI spectroscopy of high-temperature plasma for nuclear fusion provides a wealth of information which often cannot be inferred by other means \cite{tawara2003,Beiersdorfer2010}.
Hollow atoms, \ie HCIs with multiple inner holes produced by the interactions of HCI with solid surfaces, have also attracted considerable attention for years because of interests in their exotic configurations of many electrons in excited orbitals simultaneously, followed by recent studies with thin 2D materials \cite{bbce1990,btgb1996, stolterfoht2003, Niggas2024}. 
Recently, hollow atoms have been produced by x-ray free-electron lasers, and their dynamics in short timescales have been intensively studied \cite{young2010, son2011}.

Recently, x-ray microcalorimeters have been successfully applied to the HCI spectroscopy both in the laboratory and in space due to their high energy resolution, high quantum efficiency and wide spectral bandwidth \cite{beiersdorfer2003b, Beiersdorfer2009, Seely2017,kabe2017,hhaf2024}.
In astronomy, in particular, the Hitomi satellite with its x-ray microcalorimeter marked a new era by achieving high-resolution measurements of the He-like Fe$^{24+}$ transition to reveal the plasma velocity in the Perseus galaxy cluster \cite{hitomicollaboration2016}.

In this letter, we present the first experimental evidence of a new type of HCIs, namely highly charged muonic atoms with a few electrons.
When a slow negative muon encounters an atom, it is captured into a highly excited state with the emission of a bound electron and forms a muonic atom.  
The muonic atom then experiences a cascading deexcitation process by muon-induced Auger electron emission, which results in the stepwise stripping of the bound electrons of the atom, i.e., the formation of a highly charged muonic atom \cite{horvath2011}.
Since the muon is heavier than an electron (207 times), the muon orbits much closer to the nucleus than an electron would, and the muonic transition energy is again 207 times larger than the electronic one.
The cascade is followed by {\it muonic} x-ray emission and the muon is finally absorbed by the nucleus, while the intrinsic muon lifetime is 2.2~$\mu$s.
On the other hand, when bound electrons exist in their excited orbitals together with $K$-shell vacancies, {\it electronic} $K$ x rays are also emitted.
Here, we report the measurements of these electronic $K$ x rays from a few-electron bound muonic argon atom ($\mu$Ar) in the gas phase as schematically shown in Fig.~\ref{fig:scheme}.
Although such exotic atoms were predicted theoretically, detailed information on bound electrons in muonic atoms has never been obtained experimentally before. 
It is feasible to distinguish the electronic states by high-precision electronic characteristic x-ray spectroscopy.
We took full advantage of multi-pixel transition-edge-sensor (TES) superconducting x-ray microcalorimeters to achieve a high-resolution and high-detection efficiency for x-ray spectroscopy.

Recently, we performed high-precision measurements of  {\it muonic} x rays from muonic Ne atoms ($\mu$Ne) in the gas phase \cite{okumura2023}, based on our theoretical proposal to use exotic atoms as a complementary probe with respect to HCIs for bound-state strong-field QED tests \cite{paul2021}.
A pure two-body system of nuclei and a negative muon is prepared for $\mu$Ne since all bound electrons are stripped by the muon-induced Auger deexcitation process.
On the other hand, we also observed {\it electronic} characteristic $K$ x rays from muonic Fe atoms ($\mu$Fe) in a metal. The broad structured spectra revealed ultra-fast (in tens femtoseconds (fs) range) dynamics of many electrons: electron stripping by the Auger process and refilling from the surrounding Fe \cite{okumura2021b}.
Based on the $\mu$Ne and $\mu$Fe experiments, we came up with the idea; if we targeted elements with relatively large $Z$ like Ar in the gaseous target, the majority of electrons would be stripped off due to suppressed electron refilling in the gas phase, which leads to the formation of highly charged exotic ions with a few bound electrons.

To prove this idea, the experiment was carried out at the D2 beamline of the Materials and Life Science Experimental Facility (MLF) at J-PARC ~\cite{higemoto2017}.
Details of the setup were described in previous papers~\cite{okada2020,okumura2021a,okumura2021b}.
The negative muons with a momentum of 21.0~MeV/$c$ were delivered in a double-pulse structure containing $\sim$10$^4$ muons per double pulse with a repetition rate of 25~Hz.
They were injected into the gas chamber filled with 0.1 and 0.4 atm Ar gas, and emitted x rays were detected by the TES detector.

We employed a 240-pixel TES array developed by the National Institute of Standards and Technology (NIST)~\cite{doriese2017}.
The count rate of x rays on each pixel was kept to be less than a few counts per muon pulse.
For the online energy calibration, we monitored characteristic $K$ x rays of S, Cl, K, and Fe produced by an x-ray generator.
As discussed in the previous study \cite{okumura2023}, 
the energy shift due to the thermal-crosstalk effect in the pulsed-beam conditions limits the accuracy of the absolute energy, which was evaluated to be below 0.3 eV by the comparison of measured x-ray energies of the calibration characteristic $K$ x rays under the on- and off-beam conditions.
We evaluated the energy resolution $\Delta E=5.3~\textrm{eV}$ (FWHM) under the off-beam condition using the Cl $K\alpha$ peaks.
The response function of the present TES detector is accompanied by the low-energy tail.

To measure low-energy x rays below 3~keV, we employed a low-energy x-ray filter made of extremely thin polyimide and aluminum supported by a mesh (LUXEL LEX-HT) as a vacuum window, and three IR-blocking filters consisting of a 110 nm-thick Al film and a 200 nm-thick polyimide in front of the TES detector. 
The net transmission of 2.8 keV photons is about 72\%. 
We prepared the low-pressure Ar gas target as low as 0.1 and 0.4~atm at room temperature.
We corrected the intensities of x-ray spectra by considering the gas target's self-absorption, the transmission of the vacuum window and the Al filters, and the absorption efficiency of the TES absorber.

\begin{figure*}[bth]
	\begin{center}
		\includegraphics[keepaspectratio=true,,width=0.9\linewidth]{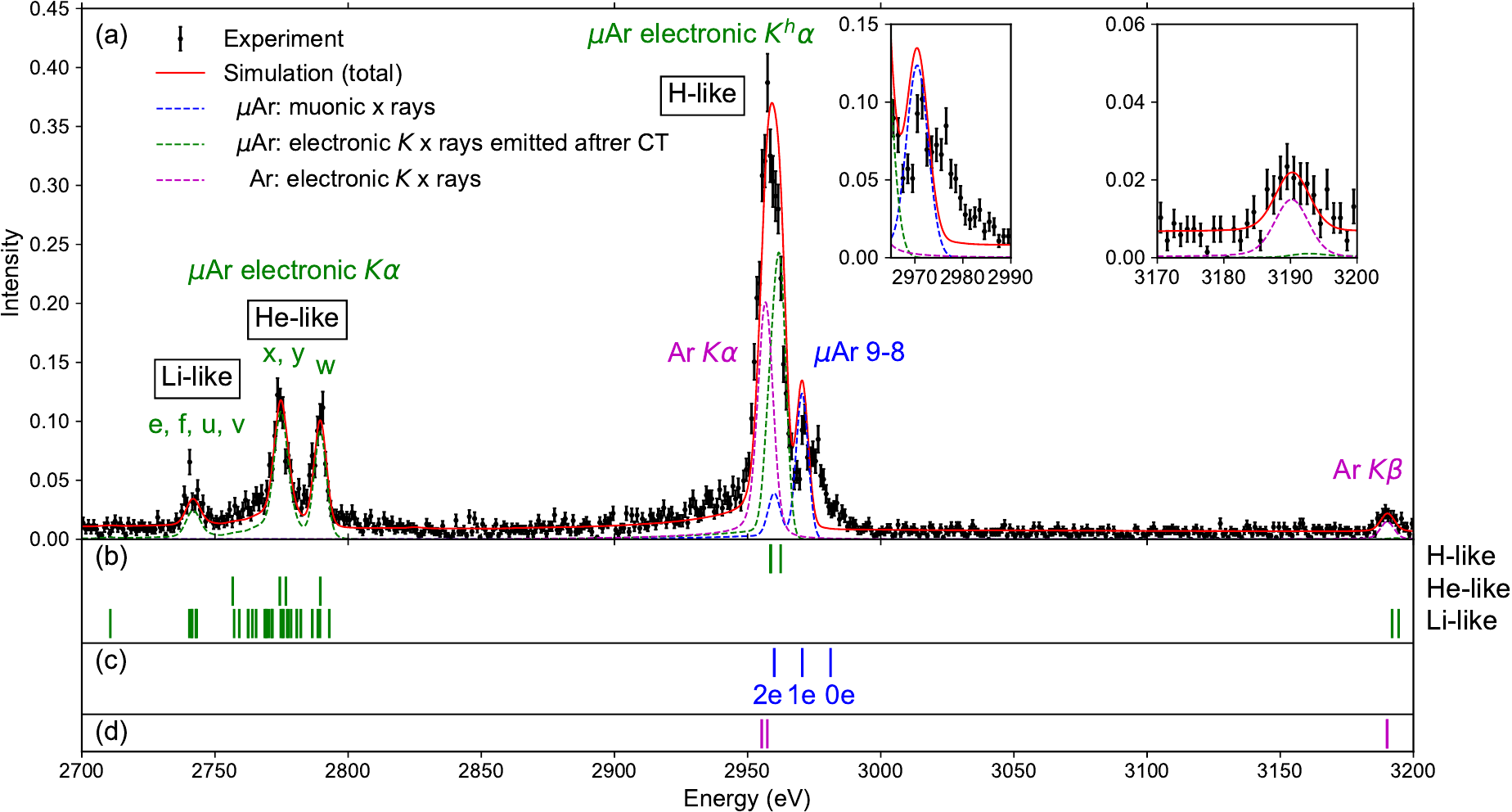}
	\end{center}%
	\caption{ \label{fig:wide_spectrum} 
(a) An observed x-ray spectrum in the range of 2700-3200~eV at the Ar pressure of 0.1~atm with the simulation. 
The inserted figures show enlarged regions of the shoulder at the high-energy side of the peak at 2960.5(3)~eV and the Ar K$\beta$ peak around 3190 eV.
The x-ray emission prior to CT is also included; However, the intensity is negligibly small.
(b) Theoretical x-ray energies for the electronic transition of highly-charged Cl$^{q+}$ ($q$=14,15,16) ions calculated by the HULLAC code \cite{bar-shalom2001}.  
(c) Theoretical muonic x ray energies of the 9-8 transition of $\mu$Ar accompanied with 0,1,2 $K$-electrons obtained by the MCDFGME code \cite{mallow1978,santos2005,trassinelli2007,indelicato2013,paul2021}. 
(d) Energies of Ar $K\alpha$ and $K\beta$ x rays \cite{deslattes2003}.
}
\end{figure*}

We obtained x-ray spectra by summing up those from all TES pixels working properly after selecting the events within a time window for the muon-beam induced signals \cite{okada2020}.
The x-ray spectrum in the energy range from 2700 to 3200~eV at a target-gas pressure of 0.1~atm is shown in Fig.~\ref{fig:wide_spectrum}(a).
We found characteristic peak structures in the two regions: one is from \SIrange[]{2720}{2820}{\electronvolt}, and the other is from \SIrange[]{2900}{3000}{\electronvolt}.
These resolved peaks mainly stem from electronic $K$ x rays emitted from $\mu$Ar, and they are clear evidence of the formation of muonic atoms with a few bound electrons.
We also observed an isolated peak at 3190.7(3)~eV. 
It is ascribed to Ar $K\beta$ x ray (3190.49(24)~eV \cite{deslattes2003}), 
which is caused by the direct $K$-shell ionization of Ar by the incident muons and possibly by energetic electrons accompanied by the main muon beam. 

The time scale of the muon deexcitation cascade of $\mu$Ar is of the order of 100~fs, and after reaching the ground state of $n_\mu=1$ ($n_\mu$: the muon principal quantum number), it is absorbed by the nucleus with the lifetime of 580~ns \cite{bertin1973}.
On the other hand, after the muon deexcitation cascade, a charge-transfer (CT) process from the surrounding Ar is inevitable because its cross section is estimated to the order of 10$^{-14}$~cm$^2$ \cite{vancura1994}.
This corresponds to the rate of $\sim$1 ns$^{-1}$ at the gas pressure of 0.1~atm, which is much faster compared to the available time range of several hundred ns before the muon's nuclear absorption.

The important aspect for understanding the dynamics involved is whether electronic $K$ x rays are emitted before or after CT.
The electron stripping by muon-induced Auger processes occurs from $M$-shell electrons down to $K$-shell ones.
From the energy conservation, Auger processes of $\Delta n_\mu=1$ accompanied by $L$-shell or $K$-shell electron emission are expected to start at $n_{\mu}=17$ or 8, respectively. 
Since most of the $L$-shell electrons are lost when a $K$-shell hole is produced, electronic $K$ x-ray emission during the muon deexcitation cascade is hardly expected.
Therefore, it is readily concluded that most of electronic $K$ x rays are emitted after CT.
This is a sharp contrast to the deexcitation process of $\mu$Fe in a metal, where $K$-shell ionization starts while several $L$-shell electrons remain \cite{okumura2021b}.
We should note that, at the moment of electronic $K$ x-ray emission after CT, the nuclear charge of Ar ($Z=18$) is completely screened by the $1s$ muon, and their electronic energy levels are comparable with those of the $Z-1$ atom, chloride (Cl).
We plotted the calculated electronic $K$ x-ray energies from highly charged Cl ions in Fig.~\ref{fig:wide_spectrum}(b).

\begin{table}[t]
	\caption{\label{tab:peaks}
	Observed peak energies, assignment of the transitions, and theoretical energies.
	Numbers in parentheses represent experimental errors.
	Labels for the transitions, such as $x,y,$ and $w$, follow Gabriel's notations \cite{gabriel1972}.
	Transitions of negligible intensity are not listed.
	The theoretical energies for electronic $K$ x rays were calculated by the HULLAC code \cite{bar-shalom2001} assuming the full screening of the nuclear charge by the muon, i.e., HCIs of Cl, while muonic x-ray energies were evaluated by the MCDFGME methods  \cite{mallow1978,santos2005,trassinelli2007,indelicato2013,paul2021}. 
 	The relative intensities obtained from the cascade simulation are also listed. (see the main text).
}
	\begin{ruledtabular}
		\begin{tabular}{dcrr}
\multicolumn{1}{l}{Observed}	&	\multicolumn{1}{c}{}	&	\multicolumn{1}{c}{Theoretical}	&	\multicolumn{1}{c}{Calculated}	\\
\multicolumn{1}{c}{energy (eV)}	&	\multicolumn{1}{c}{Assignment}	&	\multicolumn{1}{c}{energy (eV)}	&	\multicolumn{1}{c}{intensity}	\\
\hline
2741.5(3)	&	Li-like $e$	($^4P_{5/2}$ - $^2P^o_{3/2}$)$^a$	&	2742.93~~~	&	0.011	\\
	&	Li-like $f$ ($^4P_{3/2}$ - $^2P^o_{3/2}$)$^b$	&	2741.34~~~	&	0.012	\\
	&	Li-like $u$ ($^4P^o_{3/2}$ - $^2S_{1/2}$)$^c$	&	2741.38~~~	&	0.020	\\
	&	Li-like $v$	($^4P^o_{1/2}$ - $^2S_{1/2}$)$^d$	&	2740.42~~~	&	0.010	\\
2774.4(3)	&	He-like $x$ ($^3P_2$ - $^1S_0$)	&	2776.63~~~	&	0.042	\\
	&	He-like $y$ ($^3P_1$ - $^1S_0$)	&	2774.36~~~	&	0.228	\\2788.9(3)	&	He-like $w	$ ($^1P_1$ - $^1S_0$)	&	2789.55~~~	&	0.237	\\
\hline
2960.5(3)	&	H-like $2p_{1/2}$ - $1s_{1/2}$	&	2958.59~~~	&	0.292	\\
	&	H-like $2p_{3/2}$ - $1s_{1/2}$	&	2962.41~~~	&	0.582	\\
	&	Ar $K\alpha$1,2	&	2957.682$^e$	&	0.160	\\
	&	$\mu$Ar 9-8 with 2e	&	2959.8347	&	0.102	\\
2971.4(3)	&	\hspace{4 pt}with 1e	&	2970.0527	&	0.347	\\
-	&	\hspace{4pt}with 0e	&	2981.0349	&	0.000	\\
2975.6(3)	&	unassigned	&	-~~~~	&	-	\\
\hline
3190.7(3)	&	Ar $K\beta$	&	3190.49$^e$~~	&	0.011	\\
	\end{tabular}
	\end{ruledtabular}
\raggedright
$^a$ $1s2p^2(^3P)^4P_{5/2}$ - $1s^22p_{3/2}{}^2P^o_{3/2}$ \\
$^b$ $1s2p^2(^3P)^4P_{3/2}$ - $1s^22p_{3/2}{}^2P^o_{3/2}$ \\
$^c$ $1s2s2p(^3P^o)^4P^o_{3/2}$ - $1s^22s^2S_{1/2}$ \\
$^d$ $1s2s2p(^3P^o)^4P^o_{1/2}$ - $1s^22s^2S_{1/2}$ \\
$^e$ Experimental values from \cite{deslattes2003} 
\end{table}

Observed peak energies and their assignment are summarized in Table \ref{tab:peaks}.
In the lower energy region, three peaks are well-resolved.
Two peaks at 2788.9(3) and 2774.4(3)~eV are consistent to electronic $K$ x rays from He-like Cl with a single hole in the $K$-shell:  the higher corresponds to $^{1}P_{1}$-$^{1}S_{0}$ (resonance line, $w$) and the lower one to $^{3}P_{2}$-$^{1}S_{0}$ and $^{3}P_{1}$-$^{1}S_{0}$ (intercombination lines, $x$ and $y$). 
The energies of $x$ and $y$ transitions are too close to be resolved experimentally.
For the He-like system, another peak of the transition $^{3}S_{1}$-$^{1}S_{0}$ (forbidden line, $z$) at 2756.86498~eV \cite{yerokhin2019b} should, in principle, exist; however it is not observed because the transition rate is far low compared to the time scale of CT to form Li-like satellite states and
the muon's nuclear absorption, and most of $^3S_1$ states do not survive until $K$ x-ray emission.
The peak at 2741.5(3)~eV is consistent with the emission accompanied by $e, f$, and $u, v$ transitions from Li-like Cl ions \cite{yerokhin2017}, which are not well separated with the present energy resolution. 

In the higher energy region, we observed a peak at 2960.5(3)~eV with a clearly discernible shoulder component on the higher-energy side.
The peak position is consistent with the hyper-satellite electronic $K^h\alpha$ x rays associated with the $2p_{1/2}$-$1s_{1/2}$ and $2p_{3/2}$-$1s_{1/2}$ transition of H-like $\mu$Ar with two holes in the $K$ shell. 
The shoulder component originates from the {\it muonic} x-ray emission associated with the transition from $n_{\mu}=9$ to 8. 
The detailed discussion will be described later.
In addition, it is naturally expected that a component of Ar $K\alpha$ x rays (2957.682(16)~eV \cite{deslattes2003}) is contained in this peak, because we observed the isolated peak of Ar $K\beta$ x rays.

We numerically simulated the deexcitation cascade after the muon capture into the muonic state of specific principal quantum numbers $n_{\mu}$ and the subsequent CT. 
To trace the temporal evolution of electron configurations of $\mu$Ar, precise information on the transition rates and the energy levels involved in the deexcitation cascade is required.
They are calculated by the HULLAC code \cite{bar-shalom2001} assuming the full screening of the nuclear charge by the muon, i.e., Cl. 
Dynamical changes in electron binding energies during the deexcitation cascade also play key roles because they determine the timing when $K$-shell holes are produced by muon-induced Auger processes.
We calculated the muon-induced Auger rate, which depends on the electron binding energy at each step of the cascade, based on the Akylas-Vogel code \cite{akylas1978}.
In addition to the electronic $K$ x rays, we also calculated muonic x-ray energy spectra emitted during the deexcitation cascade.
They were evaluated by the MCDFGME code \cite{mallow1978,santos2005,trassinelli2007,indelicato2013,paul2021}, which accurately took into account the repulsive interaction between the muon and the electrons.
We simulate the cascade process with an initial muonic state of  $n_{\mu}$=38, $l_{\mu}$=37, where the muon is mainly captured into a circular muonic state with ($l_{\mu}=n_{\mu}-1$).
This initial condition was employed by analyzing the wavefunctions of the muon and electron by relativistic density functional theory and finding that Auger decay begins when the muon is captured to the muonic states of $n_{\mu}$=38 \cite{Tong2023}.
Note that we also predict that the Auger decay for $\mu$Ne atoms should start below $n_{\mu}$=27, which is in reasonable agreement with the estimation of the initial capture state speculated by the relative intensity ratio of the muonic x rays \cite{Kirch1999}.  
A detailed description can be found elsewhere \cite{Tong2024}.
With regards to CT, we assumed that an electron is transferred from the surrounding Ar to the electronic excited states of $n$ = 9 of $\mu$Ar based on the classical over-barrier model \cite{niehaus1986, ryufuku1980}. 
We employed the CT cross section of $1.211\times10^{14}$ cm$^2$ measured for the Ar$^{16+}$ + Ar system \cite{vancura1994}, which mimics the present system, $\mu$Ar + Ar.
The more details are described in the Supplementary Material.

Figure~\ref{fig:wide_spectrum}(a) compares the simulation results to the observed spectrum. 
After adding a constant background, the simulation results were normalized to the experimental spectrum below 2850 eV.
We evaluate the fraction of the Ar $K\alpha$ x-ray component from the $K\beta$ x-ray yield, using the $K\alpha1$/$K\beta$ x-ray intensity ratio of 0.11 reported in \cite{scofield1974}.
The simulation excellently reproduces the relative observed peak intensities and structures with the low energy tail TES response function of $\Delta E=5.3~\textrm{eV}$, where no fitting parameter was employed in the simulation for x-ray emission from $\mu$Ar. 
This justifies the validity of our model of the involved dynamics. 
We confirmed that electronic $K$ x rays emitted during the muon cascade prior to CT are significantly small (1.5\%) compared with those after CT, as expected. 
The simulation shows that electronic $K^{h}\alpha$ x rays from H-like $\mu$Ar are emitted after the first single CT, while most of the electronic $K\alpha$ x rays from He-like and Li-like $\mu$Ar are after the second and the third CT, respectively.
Thus,  CT is essential for these electronic x-ray emissions.
Concerning the $x$ and $y$ intercombination lines, the simulation predicted that the $y$ line is dominant and the $x$ line is 18\% of the $y$ line. 
It is explained by the fact that, due to the low radiative rate ($1.9\times10^{8}$~s$^{-1}$) for the $x$-line, the majority of $^3P_2$ states are lost due to further CT before emitting x rays.
It should be noted that we observed no significant target pressure dependence between 0.1 and 0.4 atm conditions.
This behavior was also reproduced by the simulation.
Regardless of the target pressure, CT always occurs within 580~ns before the muon's nuclear absorption. 

The deexcitation dynamics before CT can also be extracted from the observed muonic x-ray spectrum.
Fig.~\ref{fig:wide_spectrum}(c) represents the calculated muonic x-ray energies from 9-8 transitions under the conditions where 0, 1, or 2 $K$-shell electrons remain in $\mu$Ar.
Due to the interaction between the muon and electrons, an energy shift of 10 eV/electron is expected.
From a comparison of Figs.~\ref{fig:wide_spectrum}(a) and (c),  it is evident that the shoulder component of the hyperstellite $K^h\alpha$ x-ray peak is consistent with the 9-8 muonic x rays with 1 $K$-shell electron in terms of the transition energy and intensity, while a small component from $\mu$Ar with 2 $K$-shell electrons is included in the main peak.
These peak intensities are very sensitive to the muon deexcitation dynamics, particularly the number of $L$-shell electrons remaining in $\mu$Ar.
The 9-8 radiative transition is calculated to be slow ($4.4\times10^{12}$ s$^{-1}$) compared with muon-induced $L$-shell Auger processes.
An intense 9-8 muonic x-ray emission thus requires the condition where $L$-shell electrons are completely peeled off before the muon cascades down to $n_\mu=9$.
As already pointed out, a muon-induced $K$-shell Auger process with $\Delta n_\mu=1$ starts from $n_\mu=8$.
The observation of the peak from $\mu$Ar with 1 $K$-shell electron suggests a contribution from the additional minor process of a $K$-shell Auger emission with $\Delta n_\mu>1$ before reaching $n_\mu=9$.
The most probable transition is from $n_\mu=11$ to 9 ($4.3\times10^{12}~\textrm{s}^{-1}$), which means very few numbers of $L$-shell electrons remain already at $n_\mu=11$. 
The simulated intensities of the 9-8 muonic x-ray peaks agree with the experimental result, which again confirms the validity of the present cascade model.
In addition, we remark that muonic x rays of the 10-9 transition were clearly observed around 2100~eV. It is another evidence of the reduced number of remaining $L$-shell electrons also at $n_\mu$=10.
On the other hand, we also noticed the additional component higher than the shoulder peak in Fig.~\ref{fig:wide_spectrum}(a), which was not reproduced by our simulation.

Here, we comment on the time region of involved deexcitation dynamics between $\mu$Ne, $\mu$Ar, and $\mu$Fe.
It is also important to recognize the difference in the time scale of the muon- and electron-deexcitation, both of which depend heavily on $Z$ of the nucleus. 
As already described, for $\mu$Ne in the gas phase, the muon cascade ends within 1~ps, and the muon-induced Auger process strips all electrons off. 
For measuring muonic x rays of the 5-4 transition from $\mu$Ne \cite{okumura2023},
we emphasize that we can prepare the ideal two-body system of a muon and a nucleus when the muon stays at $n_{\mu}$=5. 
If we focus on the delayed time region, CT followed by the $K$ x-ray emission will occur, although its observation is challenging because of the low energy ($<$ 1.0 keV) of x rays. 
In $\mu$Fe in a metal,  the muon cascade ends within a few tens fs, and the electron dynamics also finish in the same time scale by forming neutral atoms where all electronic vacancies are filled by electrons from the surrounding.
In $\mu$Ar in the gas phase,  the majority of electrons are stripped off before the muon reaches the ground state within \SI{100}{\femto\second}. However, CT followed by the $K$ x-ray emission will occur and be experimentally observable in the energy region around 3~keV, as we confirmed.

In summary, we have confirmed the formation of highly charged muonic atoms, which are exotic few-body systems composed of a muon, a few electrons, and a nucleus. These systems have never been directly observed experimentally before. 
Our confirmation was achieved through measurements of characteristic x-rays in the 2.7-3.0 keV range from H-, He-, and Li-like $\mu$Ar in the gas phase. 
We utilized a high-precision TES x-ray detector with a resolution of 5.3 eV FWHM. 
Specifically, we identified two peaks at 2788.9(3) and 2774.4(3) eV as $K$ x-rays corresponding to transitions $^1P_1$-$^1S_0$ ($w$) and $^3P_1$-$^1S_0$ ($y$) / $^3P_2$-$^1S_0$ ($x$) of He-like $\mu$Ar, and a peak at 2741.5(3) eV as the $K$ x-rays corresponding to transitions $e$,$f$ and $u$,$v$ of Li-like $\mu$Ar. 
We also identified H-like $\mu$Ar through a hyper-satellite peak of $2p_{1/2}$-$1s_{1/2}$ and $2p_{3/2}$-$1s_{1/2}$ transition at 2960.5(3) eV. 
These electronic $K$ x-rays are emitted following charge transfer from the surrounding Ar, occurring after the muon's deexcitation to the ground state. 
The observed energies of these electronic $K$ x-rays align with those of chloride (Cl) ions due to the muon's full screening of the Ar nucleus charge.
Overall, our findings are in excellent agreement with our cascade simulation. 
These results provide a new approach for spectroscopy of highly charged ions and establish a platform to explore quantum dynamics, particularly focusing on the interaction between different negative particles such as a muon and electrons bound by a single nucleus. 
This approach extends beyond muonic atoms to include other exotic systems like antiprotonic atoms \cite{simons1994}.

Finally, high-precision measurements of 44~keV muonic x rays of the 4-3 transition from $\mu$Ar are now undergoing to verify the large QED effect up to 100~eV.
The contribution of possible remaining electrons is critical. 
From the present results, we can estimate that the chance of a remaining $K$-shell electron is less than 0.1\% when the muon reaches to $n_{\mu}=4$. 
The energy shift due to the 0.1\% contribution from $\mu$Ar with one $K$-shell electron is at a 1~meV level, which is smaller than the expected accuracy for the absolute energy and will not limit the precision of the QED test.

\vspace*{5mm}
\begin{acknowledgments}
The muon experiment at the J-PARC Materials and Life Science Experimental Facility was performed under a user program (2019MS01).
This work was supported by the JSPS KAKENHI (Grant-in-Aid for Scientific Research on Innovative Areas, Toward new frontiers: encounter and synergy of state-of-the-art astronomical detectors and exotic quantum beams 18H05457, 18H05458, 18H05460, 18H05461, 18H05463, and 18H05464, Grant-in-Aid for Scientific Research (A) 18H03713, 18H03714 and 23H00120, Grant-in-Aid for Scientific Research (B) 23H03660, Grant-in-Aid for Scientific Research (C) 22K03493, and Grant-in-Aid for Young Scientists 20K15238), the RIKEN Pioneering Projects, 
and the NIFS collaboration program (NIFS23KIIF031).
XMT was supported by the Multidisciplinary Cooperative Research Program in Center for Computational Sciences, University of Tsukuba.
N.P. thanks CNRS Institute of Physics for support and RIKEN for a young scientist fellowship. 
P.I. is a member of the Allianz Program of the Helmholtz Association, contract no EMMI HA-216 “Extremes of Density and Temperature: Cosmic Matter in the Laboratory".
\end{acknowledgments}

\bibliographystyle{apsrev}
\bibliography{ms.bib}
\end{document}


\title{Supplementary Material: \\
Few-electron highly charged muonic Ar atoms verified by electronic $K$ x rays}

\author{T. Okumura\orcidlink{https://orcid.org/0000-0002-3037-6573}}\email{tokumura@tmu.ac.jp}
\affiliation{Department of Chemistry, Tokyo Metropolitan University, Hachioji, Tokyo 192-0397, Japan}
\author{T. Azuma\orcidlink{https://orcid.org/0000-0002-6416-1212}}
\email{toshiyuki-azuma@riken.jp}
\affiliation{Atomic, Molecular and Optical Physics Laboratory, RIKEN, Wako 351-0198, Japan}
\author{D. A. Bennett}
\affiliation{National Institute of Standards and Technology, Boulder, Colorado 80305, USA}
\author{W. B. Doriese}
\affiliation{National Institute of Standards and Technology, Boulder, Colorado 80305, USA}
\author{M. S. Durkin} 
\affiliation{Department of Physics, University of Colorado, Boulder, CO, 80309, USA}
\author{J. W. Fowler} 
\affiliation{National Institute of Standards and Technology, Boulder, Colorado 80305, USA}
\author{J. D. Gard}
\affiliation{Department of Physics, University of Colorado, Boulder, CO, 80309, USA}
\author{T. Hashimoto}
\affiliation{RIKEN Cluster for Pioneering Research, RIKEN, Wako, 351-0198, Japan}
\affiliation{RIKEN Nishina Center, RIKEN, Wako 351-0198, Japan}
\author{R. Hayakawa}
\affiliation{International Center for Quantum-field Measurement Systems for Studies of the
Universe and Particles (QUP), High Energy Accelerator Research Organization (KEK), Tsukuba, Ibaraki 305-0801, Japan}
\author{Y. Ichinohe}
\affiliation{RIKEN Nishina Center, RIKEN, Wako 351-0198, Japan}
\author{P. Indelicato\orcidlink{https://orcid.org/0000-0003-4668-8958}}
\affiliation{Laboratoire Kastler Brossel, Sorbonne Universit\'{e}, CNRS, ENS-PSL Research University, Coll\`{e}ge de France, Case 74, 
 4, place Jussieu, 75005 Paris, France}
\author{T. Isobe}
\affiliation{RIKEN Nishina Center, RIKEN, Wako 351-0198, Japan}
\author{S. Kanda}
\affiliation{High Energy Accelerator Research Organization (KEK), Tsukuba, Ibaraki 305-0801, Japan}
\author{D. Kato}
\affiliation{National Institute for Fusion Science (NIFS), Toki, Gifu 509-5292, Japan}
\affiliation{Interdisciplinary Graduate School of Engineering Sciences, Kyushu University, Fukuoka 816-8580, Japan}
\author{M. Katsuragawa}
\affiliation{Kavli IPMU (WPI), The University of Tokyo, Kashiwa, Chiba 277-8583, Japan}
\author{N. Kawamura}
\affiliation{High Energy Accelerator Research Organization (KEK), Tsukuba, Ibaraki 305-0801, Japan}
\author{Y. Kino}
\affiliation{Department of Chemistry, Tohoku University, Sendai, Miyagi 980-8578, Japan}
\author{N. Kominato}
\affiliation{Department of Physics, Rikkyo University, Tokyo 171-8501, Japan}
\author{Y. Miyake}
\affiliation{High Energy Accelerator Research Organization (KEK), Tsukuba, Ibaraki 305-0801, Japan}
\author{K. M. Morgan}
\affiliation{National Institute of Standards and Technology, Boulder, Colorado 80305, USA}
\affiliation{Department of Physics, University of Colorado, Boulder, CO, 80309, USA}
\author{H. Noda}
\affiliation{Department of Astronomy, Tohoku University, Sendai, Miyagi 980-8578, Japan}
\author{G. C. O'Neil}
\affiliation{National Institute of Standards and Technology, Boulder, Colorado 80305, USA}
\author{S. Okada}\email{sokada@isc.chubu.ac.jp}
\affiliation{Department of Mathematical and Physical Sciences, Chubu University, Kasugai, Aichi 487-8501, Japan}
\affiliation{National Institute for Fusion Science (NIFS), Toki, Gifu 509-5292, Japan}
\author{K. Okutsu}
\affiliation{Department of Chemistry, Tohoku University, Sendai, Miyagi 980-8578, Japan}
\author{N. Paul\orcidlink{https://orcid.org/0000-0003-4469-780X}}
\affiliation{Laboratoire Kastler Brossel, Sorbonne Universit\'{e}, CNRS, ENS-PSL Research University, Coll\`{e}ge de France, Case 74, 
 4, place Jussieu, 75005 Paris, France}
\author{C. D. Reintsema}
\affiliation{National Institute of Standards and Technology, Boulder, Colorado 80305, USA}
\author{T. Sato}
\affiliation{Department of Physics, Meiji University, Kawaski, Kanagawa 214-8571, Japan}
\author{D. R. Schmidt}
\affiliation{National Institute of Standards and Technology, Boulder, Colorado 80305, USA}
\author{K. Shimomura}
\affiliation{High Energy Accelerator Research Organization (KEK), Tsukuba, Ibaraki 305-0801, Japan}
\author{P. Strasser}
\affiliation{High Energy Accelerator Research Organization (KEK), Tsukuba, Ibaraki 305-0801, Japan}
\author{D. S. Swetz}
\affiliation{National Institute of Standards and Technology, Boulder, Colorado 80305, USA}
\author{T. Takahashi\orcidlink{https://orcid.org/0000-0001-6305-3909}}
\affiliation{Kavli IPMU (WPI), The University of Tokyo, Kashiwa, Chiba 277-8583, Japan}
\author{S. Takeda}
\affiliation{Kavli IPMU (WPI), The University of Tokyo, Kashiwa, Chiba 277-8583, Japan}
\author{S. Takeshita}
\affiliation{High Energy Accelerator Research Organization (KEK), Tsukuba, Ibaraki 305-0801, Japan}
\author{M. Tampo}
\affiliation{High Energy Accelerator Research Organization (KEK), Tsukuba, Ibaraki 305-0801, Japan}
\author{H. Tatsuno}
\affiliation{Department of Physics, Tokyo Metropolitan University, Tokyo 192-0397, Japan}
\author{K. T\H{o}k\'{e}si}
\affiliation{HUN-REN Institute for Nuclear Research (ATOMKI), 4026 Debrecen, Hungary}
\author{X. M. Tong}
\affiliation{Center for Computational Sciences, University of Tsukuba, Tsukuba, Ibaraki 305-8573, Japan}
\author{Y. Toyama}
\affiliation{Center for Muon Science and Technology, Chubu University, Kasugai, Aichi, 487-8501, Japan}
\author{J. N. Ullom}
\affiliation{National Institute of Standards and Technology, Boulder, Colorado 80305, USA}
\affiliation{Department of Physics, University of Colorado, Boulder, CO, 80309, USA}
\author{S. Watanabe}
\address{Department of Space Astronomy and Astrophysics, Institute of Space and Astronautical Science (ISAS), Japan Aerospace Exploration Agency (JAXA), Sagamihara, Kanagawa 252-5210, Japan}
\author{S. Yamada}
\affiliation{Department of Physics, Rikkyo University, Tokyo 171-8501, Japan}
\author{T. Yamashita}
\affiliation{Department of Chemistry, Tohoku University, Sendai, Miyagi 980-8578, Japan}
%

\begin{abstract}
\end{abstract}

\maketitle
\section{Cascade simulation}
In the present simulation, the muon deexcitation cascade and the after-charge-transfer (CT) cascade are treated separately.
We assume the CT processes occur after the muon deexcitation to the 1s ground state.
It is justified by the fact that the muon cascade proceeds (in the 100 fs range) much faster than CT  (in the ns range).
The simulation thus comprises two parts: the muon deexcitation cascade accompanied by electron dynamics and the electronic cascade after CT with the muon in the $1s$ ground state.
We recently developed the simulation code of the muon cascade for the analysis of {\it electronic} $K$ x rays from $\mu$Fe in a metal \cite{okumura2021b}.
We have modified this simulation code to include 1) precise energy-level calculations that describe strong electron correlation effects of highly-charged $\mu$Ar, as well as the interaction between the muon and electrons,  
2) CT from the surrounding Ar, and 3) the subsequent radiative and Auger cascade of the electrons.

The energy level of the muonic atom $E_{n_\mu\ell_\mu\gamma}$ is specified by $(n_\mu,\ell_\mu,\gamma)$, where $n_\mu$ and $\ell_\mu$ are the principal and angular-momentum quantum numbers of the muon and $\gamma$ represents the electronic configuration.
The energy levels for the case of $n_\mu>1$ were calculated by a sum of three terms: a muon-orbital energy $E^\mu_{n_\mu\ell_\mu}$, an electronic level energy for full-screened ($n_\mu=1$) $\mu$Ar $E^\textrm{e}_\gamma$, and a correction term $\Delta E^{\mu\textrm{-e}}_{n_\mu\ell_\mu\gamma}$ which describes interaction between the muon and the electrons,
\begin{equation}
	E_{n_\mu\ell_\mu\gamma}=E^\mu_{n_\mu\ell_\mu}+E^\textrm{e}_\gamma+\Delta E^{\mu\textrm{-e}}_{n_\mu\ell_\mu\gamma}.
	\label{eq: level}
\end{equation}
$E^\mu_{n_\mu\ell_\mu}$ was obtained by a hydrogenic-atom solution of the Dirac equation without bound electrons.
$E^\textrm{e}_\gamma$ was calculated by the parametric potential and relativistic configuration interaction (RCI) methods using HULLAC \cite{bar-shalom2001},
assuming the $Z-1$ atom, i.e., Cl.
$\Delta E^{\mu\textrm{-e}}_{n_\mu\ell_\mu\gamma}$ was estimated by the relativistic Density Functional Theory (DFT) code, which has been recently developed for calculation of muonic atoms \cite{Tong2023}.

The deexcitaion cascade process is described by the rate equations;
\begin{eqnarray}
	\frac{d}{dt}P_{n_\mu\ell_\mu\gamma}\left(t\right) & = &  \sum_{n'_\mu\ell'_\mu\gamma'} \Gamma_{n'_\mu\ell_\mu'\gamma',n_\mu\ell_\mu\gamma}P_{n_\mu'\ell_\mu'\gamma'}\left(t\right) \nonumber \\
	&	 & - \Gamma_{n_\mu\ell_\mu\gamma}P_{n_\mu\ell_\mu\gamma}\left(t\right). 
	\label{eq: rate equations}
\end{eqnarray}
$P_{n_\mu\ell_\mu\gamma}\left(t\right)$ is a population of the muonic atom in the level of $(n_\mu,\ell_\mu,\gamma)$ at the time $t$, where $t$=0 is the time when the muon is captured by the atom.
$\Gamma_{n_\mu'\ell_\mu'\gamma',n_\mu\ell_\mu\gamma}$ is the rate for the $(n_\mu',\ell_\mu',\gamma')\to(n_\mu,\ell_\mu,\gamma)$ transition and $\Gamma_{n\ell\gamma}$ is the total decay rate of the level $(n_\mu,\ell_\mu,\gamma)$ including CT rates.
We consider three types of transitions: 
1) muon-induced processes ($n'_\mu>n_\mu$), i.e., muon-induced Auger processes and muonic x-ray emission, 
2) electron-induced processes, including electronic Auger processes  and electronic $K$ x-ray emission, 
and 3) CT from the surrounding atoms.
Rates for muon-induced processes were calculated by the Akylas-Vogel code \cite{akylas1978}, while those for electron-induced processes were evaluated by the HULLAC code assuming Cl.
For the calculation of muon-induced Auger rates, we take into account temporal changes in electron binding energies during the cascade, which are evaluated using the HULLAC code because the muon-induced Auger rate is a function of the kinetic energy of the Auger electron, i.e., the difference between the muon deexcitation energy and the binding energy.
It should be noted that these rates were evaluated by electronic and muonic wave functions calculated under the condition that their interaction was neglected.
The CT rate is estimated from the experimental CT cross section for Ar$^{16+}$ + Ar at the low collision energy (1.211$\times$10$^{-14}$ cm$^2$) \cite{vancura1994} along with the Ar pressure and its velocity at room temperature.
Although the CT cross section depends on charge states of $\mu$Ar, e.g., from 0.640$\times$10$^{-14}$ to 1.211$\times$10$^{-14}$ cm$^2$ in the case of Ar$^{q+}$ + Ar system ($8\le q\le16$) \cite{vancura1994},
that doesn't affect the present simulation result because CT is always much faster than the muon's nuclear absorption regardless of the charge state.

Muonic x-ray energies emitted during the cascade were calculated by the MCDFGME code \cite{mallow1978,santos2005,trassinelli2007,indelicato2013,paul2021}, which precisely includes the interaction between the muon and the electrons.
The results are summarized in Table I in the main text.
$K$ x-ray energies from $\mu$Ar, $E_{\gamma\to\gamma'}^{n_\mu \ell_\mu}$, was obtained by
\begin{equation}
	E_{\gamma\to\gamma'}^{n_\mu \ell_\mu}  =  
	E_{\gamma\to\gamma'} + \Delta E_{\gamma\to\gamma'}^{n_\mu \ell_\mu}, \label{eq: K x-ray energy}
\end{equation}
where $E_{\gamma\to\gamma'}$ is the corresponding $K$ x-ray energy from the $\gamma\to\gamma'$ transition of Cl obtained by the HULLAC code and $\Delta E_{\gamma\to\gamma'}^{n_\mu \ell_\mu}$ is the energy shift induced by the muon in the $(n_\mu, \ell_\mu)$ orbital evaluated by the relativistic DFT code \cite{Tong2023}.

As the initial conditions for the muon cascade simulation, we assumed that the muon is captured into a single circular orbital, i.e., $(n_\mu,\ell_\mu)=(38,37)$.

From the classical over-barrier model \cite{niehaus1986, ryufuku1980}, we assume that an electron is transferred from the neighboring Ar to the $n=9$ orbital of $\mu$Ar, which leads to the successive radiative cascade.
The $\mu$Ar population depleted via CT during the muon deexcitation cascade is given by the solutions of the rate equation (\ref{eq: rate equations}), i.e., 
\begin{equation}
	P^\textrm{CT}_{n_\mu\ell_\mu\gamma}=\Gamma^\textrm{CT}_{\gamma}\int_0^{+\infty}P_{n_\mu\ell_\mu\gamma}\left(t\right)dt.
	\label{eq: CT population}
\end{equation}
$\Gamma^\textrm{CT}_{\gamma}$ is a CT rate for the level $\gamma$.
This population was then adopted as the initial condition for the after-CT cascade simulation.
We assume charge-transfer electrons enter the $n=9$ orbitals with a statistical distribution for their angular momenta.
We considered rates for electronic Auger processes, electronic characteristic x-ray emissions, CT processes, as well as the muon's nuclear absorption, which leads to depletion of metastable states before emitting electronic x rays, such as the forbidden $z$ transition from He-like $\mu$Ar ($^3S\to$$^1S$), discussed in the main text.
The after-CT cascade simulation was continued until all $K$-shell holes were filled and no characteristic x rays were emitted.
The deexcitation of the charge-transfer electron ($n=9\to1$) is completed within 100 ps, followed by further CT processes with the time interval of the ns range on average.
We thus took into account $\mu$Ar ($n_\mu=1$) with 0-5 electrons, where only one electron is placed in a highly-excited orbital ($n>2$) and others in $n\le2$.
To consider a minor contribution from multiple CT processes, we also treat the levels in which both two electrons are excited to $n>2$ orbitals for He-like cases.

\bibliographystyle{apsrev}
\bibliography{supplement.bib}